\documentclass[12pt,tightenlines,aps,prd,twoside,onecolumn,english,superscriptaddress,
	showpacs,preprintnumbers,floats,amssymb,amsmath,amsfonts]{revtex4}

\usepackage{babel}
\usepackage{amsmath}
\usepackage{amsfonts}
\usepackage{amssymb}
\usepackage{dsfont}
\usepackage{graphicx}

\newcommand{\mbi}{m_{\tilde\beta}}
\newcommand{\mbio}{\bar{m}_{\tilde{\beta}}}
\newcommand{\fb}{f_{\tilde\beta}}
\newcommand{\bi}{{\tilde{\beta}}}
\newcommand{\eV}{\,\mathrm{eV}}

\newcommand{\MeV}{\,\mathrm{MeV}}
\newcommand{\GeV}{\,\mathrm{GeV}}

\begin{document}

\preprint{IGC-09/11-3}

\setcounter{page}{1}

\title{A solution of the strong CP problem via the Peccei--Quinn mechanism through the Nieh--Yan modified gravity and cosmological implications}

\author{\firstname{Massimiliano} \surname{Lattanzi}}
\email{lattanzi@icra.it}
\affiliation{ICRA and Dipartimento di Fisica, Universit\`{a} di Roma ``Sapienza'', Piazzale Aldo Moro 2, 00185 Roma, Italy}

\author{\firstname{Simone} \surname{Mercuri}}
\email{mercuri@gravity.psu.edu}
\affiliation{Institute for Gravitation and the Cosmos, Pennsylvania State University,\\ 104 Davey Lab, Physics Department, University Park, Pennsylvania 16802, USA}

\begin{abstract}
By identifying the recently introduced Barbero-Immirzi field with the QCD axion, the strong {\it CP} problem can be solved through the Peccei-Quinn mechanism. A specific energy scale for the Peccei-Quinn symmetry breaking is naturally predicted by this model. This provides a complete dynamical setting to evaluate the contribution of such an axion to the cold dark matter content of the Universe. Furthermore, a tight upper bound on the tensor-to-scalar ratio production of primordial gravitational waves can be fixed, representing a strong experimental test for this model.
\end{abstract}


\maketitle

According to the standard model (SM), two terms contribute to the strong {\it CP} violation. Specifically, the non-Hermitian quark mass matrix $M$ introduces a {\it CP} violation term proportional to ${\rm Arg}(\det M)$. This sums up to the vacuum angle of QCD, $\theta$, generating a {\it CP} violating interaction that depends on the parameter $\tilde{\theta}=\theta+{\rm Arg}(\det M)$. By measuring the electric dipole moment of the neutron, an extremely small upper limit can be fixed for $\tilde{\theta}$, which turns out to be smaller than $10^{-10}$. Such a small value implies an extremely precise compensation between two completely uncorrelated parameters: one associated with the global structure of the $SU(3)$ gauge group and the other related to the $SU(2)\times U(1)$ breaking symmetry. The unnaturalness of this ``fine tuning'' goes under the name of \emph{strong {\it CP} problem}. 

Peccei and Quinn proposed a dynamical mechanism to solve the strong {\it CP} problem \cite{PecQui77}. They postulated the existence in the standard model of an additional $U(1)$ axial symmetry, often denoted as $U(1)_{PQ}$. On the one hand, if this symmetry were exact, the {\it CP} violating interaction could be eliminated through a chiral rotation. 
On the other hand, we expect that the $U(1)_{PQ}$ is spontaneously broken by the chiral anomaly. Interestingly enough, the Peccei-Quinn (PQ) mechanism allows us to solve the strong {\it CP} problem, even though the $U(1)_{PQ}$ additional symmetry is not preserved by quantization.

In order to briefly describe how the PQ mechanism works, it is worth recalling that the spontaneous breaking of the $U(1)_{PQ}$ symmetry generates a (pseudo) Nambu-Goldstone boson, called \emph{axion} \cite{Wei78;Wil78}, a possible cold dark matter (CDM) candidate \cite{KT-K}. The axion interacts with matter through the following effective action \cite{Note2}
\begin{align}\label{01}\nonumber
S_{\rm Eff}&= S\left[A\right]+S_{\rm Dir}\left[\psi,\bar{\psi},A\right]+ S_{\rm matt}\left[\frac{d a}{f_a},\psi,\bar{\psi}\right]
\\
&+\frac{1}{2}\int\star d a\wedge d a
-\frac{g_s^2}{8\pi^2}\int\left(\tilde{\theta}+\frac{a}{f_{a}}\right){\rm tr} G\wedge G\,,
\end{align}
where $f_a$ denotes the scale of the $U(1)_{PQ}$ symmetry breaking. $G=d A+i g_s A\wedge A$ is the curvature 2-form associated with the $SU(3)$ valued connection 1-form $A=A^I\lambda^I$, $\lambda^K$ being the generators of the group, and $g_s$ the strong coupling constant. With the collective symbols $\psi$ and $\bar{\psi}$ we denoted fermion matter fields, interacting with the axion through a derivative coupling term, $S_{\rm matt}$.

The {\it CP} violating $\tilde{\theta}$ term combines with the anomaly-induced interaction between the axion and the gluon fields; the possible observables of the theory now depend on the effective vacuum angle $\theta(x)=\tilde{\theta}+\frac{a(x)}{f_{a}}$, sometimes referred to as the \emph{misalignment angle}. The effective interaction, $\theta(x){\rm tr} G\wedge G$, represents a nontrivial potential for the axion field, which selects a particular vacuum expectation value. In particular, the periodicity of the potential in the effective vacuum angle, $\theta(x)$, implies that it has a nontrivial minimum corresponding to $\theta(x)=0$ \cite{Pec98}, so that $\left\langle a(x) \right\rangle=-f_a\tilde{\theta}$. Consequently, the gluons effectively interact only with the physical axion $a_{\rm Phys}(x)=a(x)-\left\langle a(x) \right\rangle$, preserving the theory from the strong {\it CP} violation \cite{Note1}.

The physical features of the axion, as, e.g., its mass and the strength of its interactions with ordinary matter, strictly depend on the scale of the PQ symmetry breaking, $f_a$, which remains a completely free parameter, not fixed by the theory. The scope of this paper is to present a new model that allows us to solve the strong {\it CP} problem \emph{\`a la} Peccei-Quinn, with the remarkable advantage that the parameter $f_a$ turns out to be fixed by the theory. This provides a completely determined dynamics, so that the contribution of such an axion field to CDM can be estimated as a function of the initial misalignment angle.

Even more interesting, the model predicts the production of isocurvature fluctuations during inflation, allowing us to fix a tight upper limit to the tensor-to-scalar ratio, $r$. This represents an experimentally testable prediction that can, eventually, rule out the model.

Let us start by considering a generic space-time with torsion. The chiral rotation of the fermionic measure in the Euclidean path-integral generates, besides the usual Pontryagin class, a Nieh-Yan (NY) term \cite{NieYan82}, which diverges as the square of the regulator \cite{ChaZan97;ObuMieBud97;Soo99;SooCha99} (see also \cite{KreMie01;ChaZan01}), i.e.
\begin{align}\label{j}\nonumber
\delta\psi\delta\overline{\psi}\rightarrow \delta\psi\delta\overline{\psi}
\exp & \left\{\frac{i}{8\pi^2}\int\alpha\left[R_{a b}\wedge R^{a b}
+2 M^2\left(T_a\wedge T^a\right.\right.\right.
\\
&\left.\left.\left.-e_a\wedge e_b\wedge R^{a b}\right)\right]\frac{}{}\right\}\,.
\end{align}
Above, $M$ denotes the regulator, while $\alpha$ is the parameter of the transformation \cite{Note3}. In order to avoid the appearance of this divergence, one of us has recently proposed to introduce a field, $\beta(x)$, interacting with gravity through the Nieh-Yan density \cite{Mer09}, namely
\begin{align}\label{fa}
\nonumber S_{\rm Tot}&\left[e,\omega,\psi,\overline{\psi},\beta\right]=S_{\rm HP}\left[e,\omega\right]+S_{\rm D}\left[e,\omega,\psi,\overline{\psi}\right]
\\
&+\chi\int\beta(x)\left(T^a\wedge T_a-e_a\wedge e_b\wedge R^{a b}\right)\,,
\end{align}
where $\chi$ is a generic coupling constant with the dimension of energy. According to this proposal, we assume that (\ref{fa}) is the fundamental action for gravity and matter \cite{Note4}. 

In order to clarify some aspects related to the proposed modification, we write below the resulting semiclassical effective action \cite{Mer09,MerTav09}:
\begin{align}\label{STE}
\nonumber &S_{\rm eff}=S_{\rm HP}\left[e\right]+S[A]+S_{\rm Dir}\left[e,\psi,\overline{\psi}\right]+\frac{1}{2}\int\star d\tilde{\beta}\wedge d\tilde{\beta}
\\
&+\frac{1}{8f_{\tilde{\beta}}}\int\star J_{(A)}\wedge J_{(A)}-\frac{1}{2f_{\tilde{\beta}}}\int\star J_{(A)}\wedge d\tilde{\beta}
\\\nonumber
&-\frac{1}{8\pi^2}\int\left[\left(\tilde{\Theta}+\frac{\tilde{\beta}}{2f_{\tilde{\beta}}}\right)R
\wedge R
+\left(\tilde{\theta}+\frac{\tilde{\beta}}{2f_{\tilde{\beta}}}\right)G\wedge G\right]\,,
\end{align}
where we have defined the new field $\tilde{\beta}(x)=\sqrt{6k}\chi\beta(x)$ and introduced the constant $f_{\tilde{\beta}}=\frac{2}{\sqrt{6k}}\simeq 1.98\times 10^{18}{\rm GeV}$. An $SU(3)$ valued connection 1-form $A$, representing the strong interaction, has been considered as well, with $G$ being its curvature. In the last line, a trace over internal indexes is understood.

Some comments are now in order. The pure gravitational sector of the effective theory is reminiscent of the so-called Chern-Simons modified gravity, vastly studied in the literature (see the interesting and complete review \cite{AleYun09}). So, from an effective point of view, the modification of the gravitational action proposed in Eq. (\ref{fa}) reduces to a well-known theory of gravity, originating from string theory and featuring some interesting dynamical effects, well within the presently available experimental limits \cite{PreYun09}, thus persuading us to take it seriously. The resulting semiclassical effective theory does not depend on the free coupling constant $\chi$ and shares many common features with that postulated by Peccei and Quinn. It is worth noting that in fact, in the case of massless fermions, the full action with the Nieh-Yan modification presents an additional $U(1)_A$ symmetry, which is broken at an effective level by the interaction between the $\beta(x)$ field and the fields strength in the last line of (\ref{STE}) \cite{Note6}, exactly analogous to the $U(1)_{PQ}$ introduced above. The presence of these interaction terms reflects the existence of the chiral anomaly, which, in fact, spontaneously breaks the $U(1)_A$ symmetry at the energy scale $f_{\tilde{\beta}}$, naturally determined by the theory, in striking contrast with the Peccei-Quinn scenario, where $f_a$ is a free parameter of the theory.

Let us now assume that the system evolves in a symmetric space-time, characterized by a vanishing $R^{a b}\wedge R_{a b}$ term, as the unperturbed Friedmann-Robertson-Walker (FRW) cosmological model. In this hypothesis, comparing action (\ref{STE}) with (\ref{01}), one can appreciate the functional analogy of the two effective theories. This analogy strongly suggests to identify the field $\tilde{\beta}$ with the axion field $a$ (see also \cite{GatKetYun09} for a supersymmetric analogous identification); this is the essence of our proposal, which represents the main novelty of this model, whereas, in some previous papers \cite{Mer09,MerTav09}, the possible coexistence of the $\tilde{\beta}(x)$ field and the standard axion, $a(x)$ was addressed.

The term in the last line of (\ref{STE}) represents a nontrivial potential for the Barbero-Immirzi (BI) axion field, selecting a particular {\it CP} preserving vacuum state. So, by implementing a mechanism analogous to the Peccei-Quinn one, we can solve the strong {\it CP} problem via the BI-axion field.

As was noted above, in this model the symmetry breaking energy scale $f_{\tilde{\beta}}$ is fixed by the theory, allowing us to estimate the expected zero-temperature mass of the BI-axion field, which, as for the standard axion, is generated by instantonic effects \cite{Pec98}. We obtain,
\begin{equation}
^{0}\!m_{\tilde{\beta}}=\frac{f_{\pi}}{f_{\tilde{\beta}}}m_{\pi}\frac{\sqrt{m_u m_d}}{m_u+m_d}\simeq 3.04\times 10^{-12}\ {\rm eV}\,,
\label{eq:axmass}
\end{equation}
where we used the value of the pion decay constant, $f_{\pi}=93\ {\rm MeV}$, measured in the decay process $\pi^+\to\mu^++\nu_{\mu}$. As is well known, instantonic effects depend on the temperature, in particular, we expect that the greater the temperature, the smaller the mass of the BI-axion field becomes; according to the standard literature, we have that \cite{Gro80}
\begin{equation}
\mbi(T) = 
\left\{
\begin{array}{ll}
^{0}\!\mbio b \left(\frac{\Lambda}{T}\right)^4 & T \gtrsim \Lambda\,, \\[0.2cm]
^{0}\!\mbio & T \lesssim \Lambda\,, \\
\end{array}
\right.
\label{eq:mT}
\end{equation}
where we have assumed $b=0.018$ and a color anomaly index equal to $1$. $\Lambda\simeq 200 \MeV$ is the QCD scale.

So, in this model, the physical parameters, namely the mass of the field and the magnitude of its interaction with matter, are fixed by the theory. Remarkably, this allows us to reduce the parameter space of the theory and extract strong predictions from the cosmological scenario we are going to study.

In general, when dealing with axion scenarios, there are two possibilities. The
first one is that the PQ symmetry is restored after inflation, and then broken again after the Universe cools
down. This happens if the reheating temperature $T_{RH}$ is larger than the energy scale at which
the symmetry is broken. The second possibility is that the PQ symmetry is broken
during inflation and never restored afterward. In order for the symmetry to be broken during inflation,
the scale $f_{PQ}$ has to be larger than the Gibbson-Hawking temperature
$T_{GH} = H_I/2\pi$ 
associated with the cosmological horizon (here $H_I$ is the value of the Hubble expansion rate
during inflation) \cite{Lyt92}; furthermore, in order for the symmetry to stay broken after inflation,
the reheating temperature has to be smaller than $f_{PQ}$. The inflationary expansion rate is constrained by the WMAP observations \cite{Kom09} to be
$H_I\le 6.29 \times 10^{14} \GeV$. The reheating temperature
is poorly constrained by the observations and could be anywhere in the range $1\MeV - 10^{16}\GeV$. 
However in the context of the model presented here, 
the relevant energy scale $\fb \sim 10^{18}\GeV$ is so large that we will always have to deal
with the second scenario, i.e., the symmetry remains broken after inflation. Nevertheless,
we will also briefly take into account the possibility that inflation never occurred.

The fact that the PQ symmetry stays broken after the end of inflation, has two important consequences. The first is that
the initial misalignment angle of the BI field is practically constant within the region corresponding to our present
horizon, and can take any value between $-\pi$ and $\pi$. The second is that isocurvature perturbations
are produced in the BI field.

The cosmological limits on axion properties have been recently reassessed in light of the 5-year WMAP data  \cite{VG-H}. Here we will do the same for the BI axion. Axions can be produced in the early Universe through two distinct mechanisms \cite{Note7}, namely coherent production due to the initial misalignment of the axion field, 
and the decay of axionic strings. The latter is relevant only if the symmetry breaking happens after the end of inflation,
then, here, we will be concerned only with the misalignment production.
The basic idea is that, when the axion field is created, the initial value $\theta_i$ of the misalignment angle $\theta$  is displaced
from zero, since no preferred value of $\theta$ exists. 

Since the axion field is created during inflation, our present Hubble volume corresponds
to a small patch at the time of creation, where the value of $\theta_i$ can be assumed to be spatially constant.

In a flat FRW Universe, the zero mode of the dynamical field $\theta(x)$ evolves according to:
\begin{equation}
\ddot \theta + 3 H \dot \theta + \frac{1}{\fb^2}\frac{\partial V(\theta)}{\partial\theta} = 0,
\label{eq:KG}
\end{equation}
where a dot denotes the derivative with respect to cosmological time, $H$ is the Hubble parameter,
and the potential 
$V(\theta) = \mbi^2(T) \fb^2 (1-\cos\theta)$ \cite{Note:V}.
It is clear from Eq. (\ref{eq:mT}) that in the high temperature limit $T\gg \Lambda$ the
BI-axion is effectively massless. Then $V(\theta) = 0$ and $\theta=\mathrm{const}$ is a 
solution of the equation of motion and the misalignment field is frozen to its initial value, $\theta_i$, until the mass becomes
comparable to the expansion rate of the Universe, i.e. $H\sim T$, and
the field starts oscillating around $\theta = 0$. For the value of the mass considered here, this happens
at $T\simeq 52\MeV$.
We have numerically integrated the Klein-Gordon Eq. (\ref{eq:KG})
down to a temperature well below the onset of oscillations
and used entropy conservation to obtain the present number density. In the limit of small $\theta_i$, this procedure yields
\begin{equation}
n_\bi(T_0)\simeq 2.8 \times 10^{22} \theta_i^ 2 \frac{\mathrm{axions}}{\mathrm{cm}^3},
\end{equation}
corresponding to an energy density
$\rho_\bi(T_0) \simeq 85 \,\theta_i^2 \GeV /\mathrm{cm}^{3}$.
What is remarkable about this result is that, since the energy scale at which the symmetry breaking occurs 
is fixed by the theory, the present day energy density of the BI axion depends only on the initial misalignment angle.
Given that the present critical density of the Universe is $\rho_c \sim 10^{-5} \GeV/\mathrm{cm}^3$, the
above formula points to the necessity of having $\theta_i \ll 1$. In particular, we know from the recent measurements 
of the WMAP satellite \cite{Kom09} that the present dark matter density is
$\Omega_{\mathrm{dm}} h^2 = 0.1131\pm 0.0034$ at 68\% CL,
where $\Omega_{\mathrm{dm}} \equiv \rho_{\mathrm{dm}}/\rho_c$ is the density in units of the critical density, and
$h$ is the Hubble parameter in units of 100 km sec$^{-1}$ Mpc$^{-1}$. Then assuming that
BI-axions make up all the dark matter ($\Omega_\bi = \Omega_{\mathrm{dm}}$) the initial misalignment angle has to be very small:
$\theta_i \simeq 1.2\times 10^{-4}$.
Larger values of the initial misalignment angle lead to a present axion density too large with respect
to the WMAP value, so, in general, one should require that $\theta_i\le 1.2\times 10^{-4}$. 
The axion density would be diluted, and then the limits on $\theta_i$ relaxed, in the presence of a significant entropy production at a temperature below the QCD scale. 
This could be the case if the reheating temperature $T_{RH}<\Lambda$ \cite{Giu01}.

Let us also examine the possibility that inflation never occurred \cite{Note:I}. 
In this case, the initial misalignment angle would be a function of the spatial coordinates and should be replaced
with its average over $[-\pi,\pi]$, i.e. $\langle \theta_i^2\rangle=\pi^2/3$. Consequently, the present axion energy density would be completely fixed, leading to a density parameter
$\Omega_\bi h^2 \sim 10^8$, clearly overshooting the observed value by 9 orders of magnitude \cite{Note:AS}.

\begin{figure}
\includegraphics[width=\linewidth,keepaspectratio,clip]{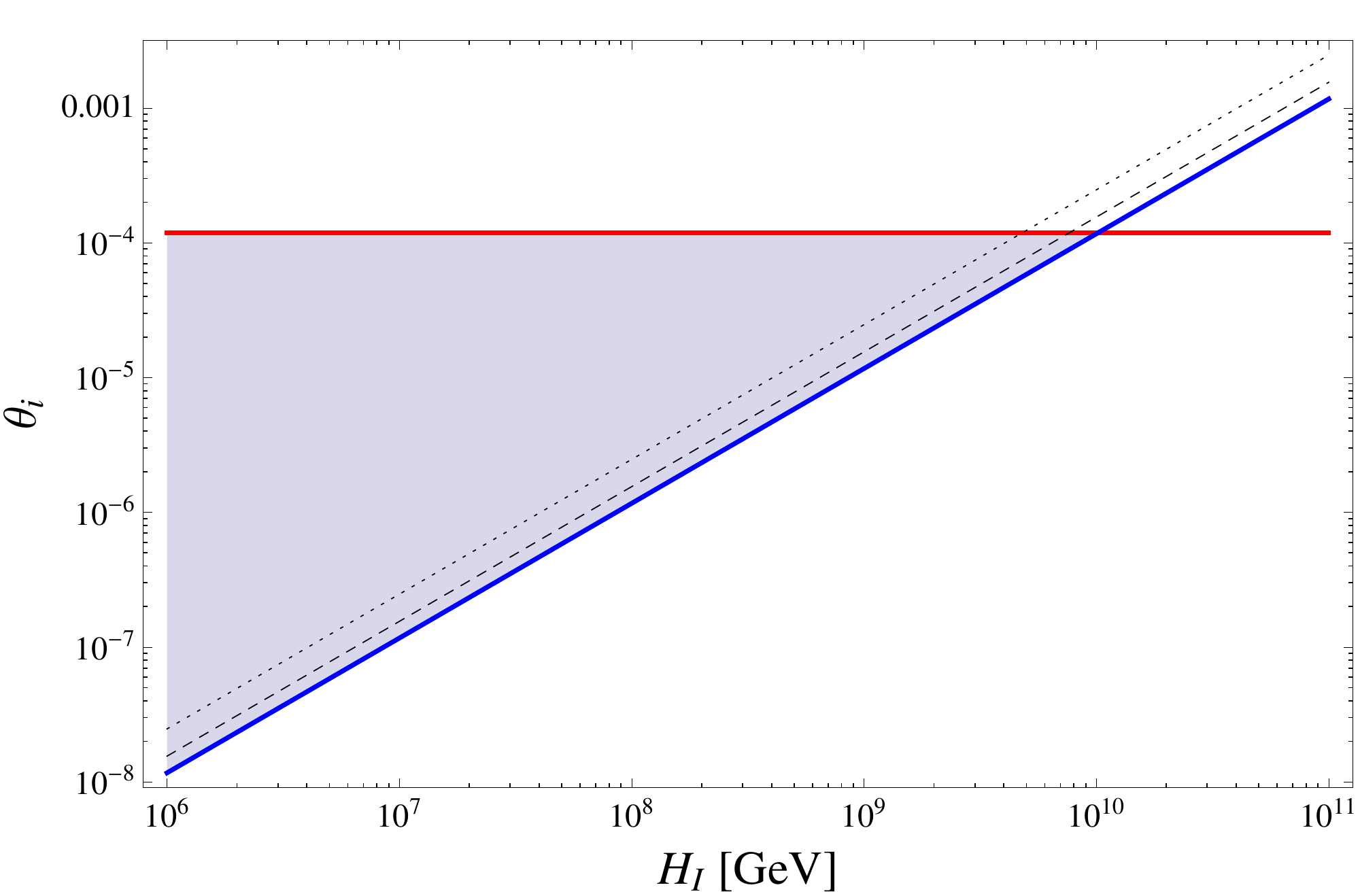}
\caption{Constraints for the BI-axion in the $(H_I,\,\theta_i)$ plane. The horizontal line corresponds to the BI-axion making all the
dark matter in the Universe (namely $\theta_i \simeq 1.2\times 10^{-4}$). The diagonal line comes from the constraints on the isocurvature fluctuations ($H_I/\theta_i <8.25 \times 10^{13}\GeV$). 
The shaded area shows the allowed parameter region. The dashed and dotted diagonal lines corresponds to the expected improvement of
the bound on isocurvature fluctuations from the measurements of the Planck satellite and from an ideal, cosmic variance limited CMB experiment, respectively 
(see \cite{VG-H} for details).}
\label{fig:pars}
\end{figure}

Another prediction of the model examined here is the production of axion isocurvature perturbations. 
As it happens for the inflaton field, de Sitter-induced quantum fluctuations in the BI-axion field are generated during inflation. 
The corresponding energy density fluctuations have an amplitude proportional to $H_I/(\theta_i f_\bi)$ and are completely uncorrelated to those in the other components (radiation and matter)
since axions were not in thermal equilibrium with photons during inflation. Moreover, since axions made a
negligible contribution to the energy budget of the Universe at that time, the fluctuations in their energy density
did not produce a corresponding perturbation in the curvature, hence the name ``isocurvature''.
The amplitude of primordial isocurvature perturbations can be constrained by observations of the CMB anisotropies;
in fact, an analysis of the WMAP data yields the constraint 
$H_I/\theta_i <4.3 \times 10^{-5} \fb =8.25 \times 10^{13}\GeV$ \cite{VG-H}.
This bound can be combined with the constraint $\theta_i <1.2 \times 10^{-4}$ to obtain the allowed region in the $(H_I,\,\theta_i)$ parameter space as shown in Fig. \ref{fig:pars}. 
In particular, the two constraints together imply $H_I \lesssim 10^{10}\GeV$. This low value 
leads to an interesting prediction of the model. Since the amount of gravitational waves
(corresponding to tensor perturbation modes) produced during inflation is also proportional to $H_I$, the model yields an upper bound
to the amplitude of tensor modes. In particular, in terms of the tensor-to-scalar ratio $r$, we get the very tight upper bound
$r<1.4\times 10^{-9}$.
This means that a detection of
even a very small amount of primordial gravitational waves by one of the upcoming
CMB experiments would rule out the model proposed here. This would be the case, in particular,
if tensor modes are detected by Planck, since it is expected to be sensitive to $r\gtrsim 0.05$.

Finally, let us briefly discuss the astrophysical constraints on the BI axion. In the case of the standard axion, the astrophysical constraints mainly depend
on the strength of its interaction with ordinary matter, since this controls the rate at which the nuclear energy generated in 
the core of a star is carried away in the form of axions, thus modifying the standard stellar evolution. In order to evade these constraints,
the axion couplings have to be either small enough so that few axions are produced as a by-product of the nuclear reactions, or large
enough in order to keep the mean free path of axions well inside the radius of the star. However, the computation of the couplings of the axion
is completely general, so that the standard results also hold for the BI axion. In particular, one has that
the couplings are inversely proportional to $f_\bi$, so that we can expect them to be extremely small
\footnote{A remarkable difference with respect to the PQ model lies in the fact that while, on the one hand, 
the $U(1)_{PQ}$ charges of the SM particles are not given by the theory, and thus have to be experimentally measured, on the other hand the Nieh-Yan ``charges''
$X_i^{(NY)}$ associated with the $U(1)_A$ symmetry of our model can
be calculated exactly from the theory. In particular, as it can be inferred from the effective action (\ref{STE}), the charges of the fundamental fermionic fields (leptons and quarks) all turn out to be equal to unity.
This universality is a direct consequence of the geometrical nature of the interaction. We would also like to remark that this completely fixes the BI-axion couplings $g_{aii}$, since these 
depend, other than from the $X_i$'s and from known quantities, only on the BI axion mass.}.
In fact, this implies that the BI-axion is, as long as astrophysical limits are concerned,
roughly equivalent to a DFSZ axion with a mass $m_a\simeq 3\times 10^{-12}\eV$. The most stringent astrophysical upper limit on the axion mass comes from the observations of
SN 1987A and states that $m_a \lesssim 10^{-3}\eV$ (or, equivalently, $f_a\gtrsim 6\times 10^9 \GeV$). Thus we conclude that astrophysical observations cannot
rule out the existence of a BI-axion.

\paragraph{Acknowledgments.}
SM is partially supported by the NSF Grant No. PHY0854743, the George A. and Margaret M. Downsbrough Endowment and the Eberly research funds of Penn State.


\begin{thebibliography}{xx}

\bibitem{PecQui77}
R.D. Peccei and H.R. Quinn, Phys. Rev. Lett. \textbf{38}, 1440 (1977); Phys. Rev. D \textbf{16}, 1791 (1977).

\bibitem{Wei78;Wil78}
S. Weinberg, Phys. Rev. Lett. \textbf{40}, 223 (1978); F. Wilczek, Phys. Rev. Lett. \textbf{40}, 271 (1978).

\bibitem{KT-K}
E.W. Kolb and M.S. Turner, \emph{The Early Universe} (Addison-Wesley, New York, 1994); M.Yu. Khlopov, \emph{Cosmoparticle Physics} (World Scientific, Singapore, 1999).

\bibitem{Note2}
The signature throughout the paper is $(+,-,-,-)$ with $\epsilon_{0123}=1$. For convenience, we set $\hbar=c=k_B=1$ and $8\pi G=k$.

\bibitem{Pec98}
R.D. Peccei, \texttt{arXiv:hep-ph/9807516}.

\bibitem{Note1}
It is worth remarking that, even postulating the existence of an additional $U(1)_{PQ}$ chiral symmetry, the electroweak {\it CP} violating effects prevent $\tilde{\theta}$ from vanishing exactly. Nevertheless, these electroweak effects induce a {\it CP} violation well within the experimental limit discussed above, i.e. $\tilde{\theta}_{\rm ew}\approx 10^{-15}$.

\bibitem{NieYan82}
H.T. Nieh and M.L. Yan, J. Math. Phys. \textbf{23}, 373 (1982).

\bibitem{ChaZan97;ObuMieBud97;Soo99;SooCha99}
O. Chand\'{i}a and J. Zanelli, Phys. Rev. D \textbf{55}, 7580 (1997); Y. Obukhov {\it et al.}, Fund. Phys. \textbf{27}, 1221 (1997); C. Soo, Phys. Rev. D \textbf{59}, 045006 (1999).

\bibitem{KreMie01;ChaZan01}
D.J. Kreimer and E.W. Mielke, Phys. Rev. D \textbf{63}, 048501 (2001); O. Chand\'{i}a and J. Zanelli, Phys. Rev. D \textbf{63}, 048502 (2001).

\bibitem{Note3}
The imaginary unit $i$ disappears in Minkowski space.

\bibitem{Mer09}
S. Mercuri, Phys. Rev. Lett. \textbf{103}, 081302 (2009).

\bibitem{Note4}
The field $\beta$ is usually referred to as the Barbero-Immirzi (BI) field; see, V. Taveras and N. Yunes, Phys. Rev. D \textbf{78}, 064070 (2008); G. Calcagni and S. Mercuri, Phys. Rev. D \textbf{79}, 084004 (2009) (see also R.~G.~Leigh, N.~N.~Hoang and A.~C.~Petkou
  JHEP {\bf 0903}, 033 (2009)
  and F. Cianfrani and G. Montani, Phys. Rev. D \textbf{80}, 084040 (2009) for different approaches).

\bibitem{MerTav09}
S. Mercuri and V. Taveras, Phys. Rev. D \textbf{80}, 104007 (2009).

\bibitem{AleYun09}
S. Alexander and N. Yunes, Phys. Rept. \textbf{480}, 1 (2009).

\bibitem{PreYun09}
N. Yunes and F. Pretorius, Phys. Rev. D \textbf{79}, 084043 (2009).

\bibitem{Note6}
It is worth remarking that the BI field turns out to be a pseudoscalar as suggested by its contribution to the irreducible torsion components \cite{Mer09} and confirmed by its equations of motion \cite{MerTav09}.

\bibitem{GatKetYun09}
S.J. Gates, S.V. Ketov, and N. Yunes, Phys. Rev. D \textbf{80}, 065003 (2009). 

\bibitem{Gro80}
D.J. Gross, R.D. Pisarski, and L.G. Yaffe, Rev. Mod. Phys. \textbf{53}, 43 (1981);
M.S. Turner, Phys. Rev. D \textbf{33}, 889 (1986).

\bibitem{Lyt92}
D.H. Lyth and E.D. Stewart, Phys. Rev. D \textbf{46}, 532 (1992).

\bibitem{Kom09}
E. Komatsu {\it et al.}  [WMAP Collaboration], Astrophys. J. Suppl. \textbf{180}, 330 (2009).

\bibitem{VG-H}
L. Visinelli and P. Gondolo, Phys. Rev. D \textbf{80}, 035024 (2009); J. Hamann {\it et al.}
JCAP 06 (2009) 022.

\bibitem{Note7}
Thermal production is excluded by astrophysical constraints.

\bibitem{Note:V}
According to the standard literature \cite{Gro80}, the form of the potential is motivated by the expected periodicity in the misalignment angle, once the gravitational instantons, depressed by the symmetries, have been neglected.

\bibitem{Giu01}
G.F. Giudice, E.W. Kolb, and A. Riotto, Phys. Rev. D \textbf{64}, 023508 (2001).

\bibitem{Note:I}
Although this is unlikely, it cannot be ruled out since other possible scenarios can avoid the shortcomings of the standard model and generate the primordial fluctuations.

\bibitem{Note:AS}
One should take into account the axion production via the decay of axionic strings as well, but this would only make our point stronger.

\end{thebibliography}
\end{document}